\documentclass{article}
\usepackage{spconf,amsmath}
\usepackage{graphicx}

\usepackage{amssymb,bm,cite,subfig}
\usepackage{algorithmic,algorithm}

\newcommand{\minimize}{\mathop{\rm minimize}\limits}

\title{LOCALIZING ACOUSTIC ENERGY IN SOUND FIELD SYNTHESIS\\BY DIRECTIONALLY WEIGHTED EXTERIOR RADIATION SUPPRESSION}
\name{Yoshihide Tomita$^1$, Shoichi Koyama$^2$, and Hiroshi Saruwatari$^1$}
\address{$^1$The University of Tokyo, 7-3-1 Hongo, Bunkyo-ku, Tokyo 113-8656, Japan\\
$^2$National Institute of Informatics, 2-1-2 Hitotsubashi, Chiyoda-ku, Tokyo 101-8430, Japan}
\begin{document}
\ninept
\maketitle
\begin{abstract}
A method for synthesizing the desired sound field while suppressing the exterior radiation power with directional weighting is proposed. The exterior radiation from the loudspeakers in sound field synthesis systems can be problematic in practical situations. Although several methods to suppress the exterior radiation have been proposed, suppression in all outward directions is generally difficult, especially when the number of loudspeakers is not sufficiently large. We propose the directionally weighted exterior radiation representation to prioritize the suppression directions by incorporating it into the optimization problem of sound field synthesis. By using the proposed representation, the exterior radiation in the prioritized directions can be significantly reduced while maintaining high interior synthesis accuracy, owing to the relaxed constraint on the exterior radiation. Its performance is evaluated with the application of the proposed representation to amplitude matching in numerical experiments.
\end{abstract}
\section{Introduction}
\label{sec:intro}

Sound field synthesis/reproduction/control aims at synthesizing the desired spatial sound inside a target region using multiple loudspeakers (or secondary sources). One or more finite interior regions of the target are surrounded by secondary sources in typical sound field synthesis systems. 

A large number of sound field synthesis techniques have been proposed for various applications. In spatial audio for virtual/augmented reality, the desired pressure (i.e., magnitude and phase) distribution is synthesized in the target region. In contrast to methods based on analytical representations derived from boundary integral equations, such as \textit{wave field synthesis} and \textit{higher-order ambisonics}~\cite{Berkhout:JASA_J_1993,Spors:AES124conv,Daniel:AES114conv,Poletti:J_AES_2005,Ahrens:Acustica2008,Wu:IEEE_J_ASLP2009,Koyama:IEEE_J_ASLP2013}, methods based on the minimization of the squared error between synthesized and desired sound fields, such as \textit{pressure matching} and \textit{mode matching}~\cite{Kirkeby:JASA_J_1993,Poletti:J_AES_2005,Ueno:IEEE_ACM_J_ASLP2019,Koyama:JAES2023}, have practical advantages since the array geometry of the secondary sources can be arbitrary. These methods can also be extended to spatial active noise control by combining with sound field estimation methods~\cite{Koyama:IEEE_ACM_J_ASLP2021,Arikawa:ICASSP2022}. In multizone sound field control for personal audio systems~\cite{Betlehem:IEEE_M_SP2015,Wu:IEEE_J_ASLP2010,Choi:JASA_J_2002,Abe:IEEE_ACM_J_ASLP2023}, the desired sound field is sometimes set using the magnitude distribution or acoustic potential energy since the desired propagation direction of the synthesized sound can be unspecified. 

Most sound field synthesis methods described above are based on solving the minimization problem of synthesis error inside the target region; however, the exterior region of the target is usually not taken into consideration. Thus, the exterior radiation power of the secondary sources can be significantly large. Several techniques to synthesize the desired sound field in the target region while suppressing the exterior radiation power, thus localizing the acoustic energy, have been proposed~\cite{Poletti:JASA2012,Ueno:IEEE_ACM_J_ASLP2019,Ueno:ICASSP2018,Arikawa:ICA2022,Kojima:EUSIPCO2023}. Basically, these methods are based on imposing a constraint on the exterior radiation power by a penalty term or (in)equality constraint in addition to the cost function of the interior synthesis error. However, suppressing the exterior radiation in all the outward directions is very difficult particularly when the number of secondary sources is not sufficiently large. An excessive constraint on the exterior radiation can also affect the synthesis accuracy in the interior region. 

In some practical situations, allowable directions of exterior radiation can be given in advance, e.g., directions where there are no people. To exploit this prior information on the priority directions of the suppression and relax the constraint on the exterior radiation power, we propose the directionally weighted exterior radiation suppression in sound field synthesis. Since the proposed directionally weighted exterior radiation power is represented by a quadratic form of the driving signal of the secondary sources, it is simply applicable to various sound field synthesis methods. We also demonstrate its application to \textit{amplitude matching}, i.e., a method to synthesize the desired amplitude (or magnitude) distribution over the target region, and its performance evaluation by numerical experiments. 

\section{Sound Field Synthesis With Exterior Radiation Suppression}
\label{sec:sfs}

\begin{figure}[t]
 \centering
 \includegraphics[width=0.6\columnwidth]{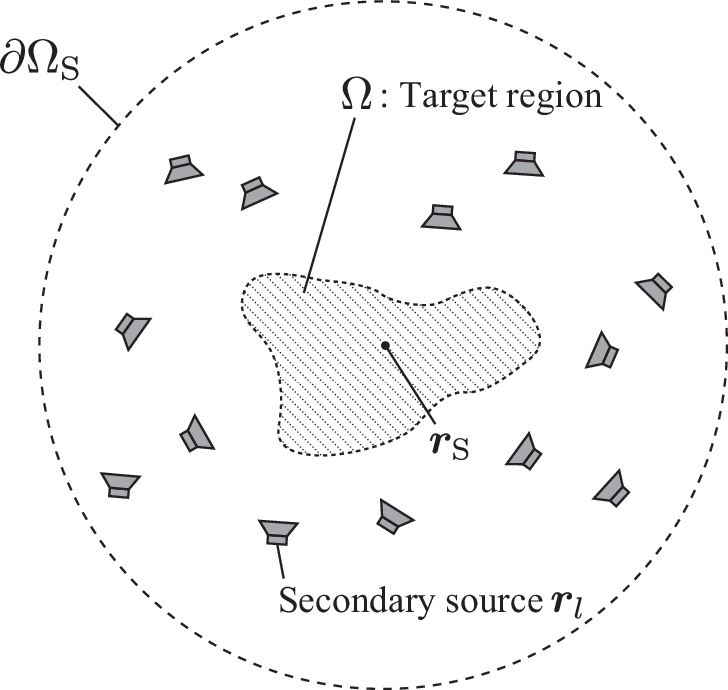}
   \caption{Sound field synthesis using multiple secondary sources.}
   \label{fig:sfc}
\end{figure}

Suppose that the desired sound field is synthesized in a target region $\Omega\subset\mathbb{R}^3$ by $L$ secondary sources at $\{ \bm{r}_l \}_{l=1}^L$. All the secondary sources are assumed to exist inside a spherical region $\Omega_{\mathrm{S}}$ ($\supset \Omega$), with the center $\bm{r}_{\mathrm{S}}$ (coordinate origin) and radius $R_{\mathrm{S}}$, and outside $\Omega$, i.e., $\bm{r}_l\in\Omega_{\mathrm{S}}\backslash\Omega$ (see Fig.~\ref{fig:sfc}). The synthesized sound field $u_{\mathrm{syn}}$ at the position $\bm{r}$ and angular frequency $\omega$ is represented by a linear combination of the transfer functions of secondary sources $\{g_l(\bm{r},\omega)\}_{l=1}^L$ as
\begin{align}
 u_{\mathrm{syn}} (\bm{r}, \omega) = \sum_{l=1}^L d_l(\omega) g_l(\bm{r},\omega), 
\end{align}
where $d_l \in \mathbb{C}$ is the driving signal of $l$th secondary source. We also define a vector of the driving signals $\bm{d}=[d_1,\ldots,d_L]^{\mathsf{T}}\in\mathbb{C}^L$. Hereafter, $\omega$ is omitted for notational simplicity.

Synthesizing the desired sound field inside $\Omega$ while suppressing the exterior radiation power is basically achieved by solving the following optimization problem:
\begin{align}
 \minimize_{\bm{d}\in\mathbb{C}^L} \mathcal{J}(\bm{d}) + \gamma \mathcal{E}(\bm{d}) + \alpha \mathcal{R}(\bm{d}),
\label{eq:cost_gnrl}
\end{align}
where $\mathcal{J}(\bm{d})$ evaluates the synthesis error inside $\Omega$, $\mathcal{E}(\bm{d})$ is a measure for the exterior radiation power of the secondary sources, $\mathcal{R}(\bm{d})$ is the regularization term for $\bm{d}$ (typically, $\mathcal{R}(\bm{d}):=\|\bm{d}\|^2$), and $\gamma$ and $\alpha$ are positive constants. In (weighted) pressure matching, $\mathcal{J}(\bm{d})$ is formulated as the squared synthesis error of pressure at control points distributed over $\Omega$~\cite{Kirkeby:JASA_J_1993,Koyama:JAES2023}. $\mathcal{J}(\bm{d})$ can also be represented by using spherical wavefunction expansion coefficients of $\{g_l(\bm{r})\}_{l=1}^L$ and the desired sound field as in (weighted) mode matching~\cite{Poletti:J_AES_2005,Ueno:IEEE_ACM_J_ASLP2019}. In amplitude matching~\cite{Abe:IEEE_ACM_J_ASLP2023}, squared synthesis error of magnitude at the control points is used for $\mathcal{J}(\bm{d})$ for multizone sound field control. Although \eqref{eq:cost_gnrl} is defined by adding $\mathcal{E}(\bm{d})$ as a penalty term, it is also possible to impose an (in)equality constraint on $\mathcal{E}(\bm{d})$~\cite{Arikawa:ICA2022,Kojima:EUSIPCO2023}. Another approach is to maximize the ratio between interior and exterior acoustic potential energies as in acoustic contrast control~\cite{Choi:JASA_J_2002}. In any case, it is necessary to appropriately define the exterior radiation power $\mathcal{E}(\bm{d})$ as a simple function of $\bm{d}$ as possible. 

\section{Directionally Weighted Exterior Radiation Suppression}

In \cite{Ueno:ICASSP2018}, the exterior radiation power $\mathcal{E}(\bm{d})$ is formulated as an angular integral of the acoustic intensity in the outward normal direction. However, it is sometimes difficult to suppress exterior radiation power in all directions without sufficient secondary sources. To enable relaxing the constraint on the exterior radiation power by using prior knowledge on allowable directions of exterior radiation, we formulate directionally weighted exterior radiation power. 

\subsection{Formulation of Directionally Weighted Exterior Radiation Power}

The exterior radiation power from $\partial \Omega_{\mathrm{S}}$, i.e., the boundary of $\Omega_{\mathrm{S}}$, is defined as 
\begin{align}
 \mathcal{E}_{\mathrm{uni}}(\bm{d}) = \int_{\partial \Omega_{\mathrm{S}}} I_r(\bm{r}) \mathrm{d} \bm{r},
\label{eq:erp_uni}
\end{align}
where  $I_r(\bm{r})$ is the acoustic intensity in the outward normal direction of $\partial \Omega_{\mathrm{S}}$, represented by $r$: 
\begin{align}
 I_r(\bm{r}) = \frac{1}{2} \mathrm{Re} \left[ u_{\mathrm{syn}}(\bm{r}) \frac{\mathrm{j}}{\rho c k} \frac{\partial}{\partial r} u_{\mathrm{syn}}(\bm{r})^{\ast} \right].
\end{align}
Here, $c$ is the sound speed, $k:=\omega/c$ is the wave number, $\rho$ is the density of air, $\mathrm{j}$ is the imaginary unit, and $\mathrm{Re}[\cdot]$ represents the real part of a complex number. We refer to $\mathcal{E}_{\mathrm{uni}}(\bm{d})$ defined in \eqref{eq:erp_uni} as uniformly weighted exterior radiation power. In \cite{Ueno:ICASSP2018}, it is shown that $\mathcal{E}_{\mathrm{uni}}(\bm{d})$ is represented by a quadratic form of $\bm{d}$ with a matrix consisting of spherical wavefunction expansion coefficients of $\{g_l(\bm{r})\}_{l=1}^L$ and translation operator. 

We define directionally weighted exterior radiation power $\mathcal{E}_{\mathrm{dir}}(\bm{d})$ as
\begin{align}
 \mathcal{E}_{\mathrm{dir}}(\bm{d}) = \int_{\partial \Omega_{\mathrm{S}}} I_r(\bm{r}) w(\theta,\phi) \mathrm{d} \bm{r},
\label{eq:erp_dir}
\end{align}
where $w: \mathbb{S}_2 \to \mathbb{R}_{\ge 0}$ is the directional weighting function, and $\theta$ and $\phi$ are respectively zenith and azimuth angles. By using $w$, one can design a directional priority of exterior radiation penalized in the cost function \eqref{eq:cost_gnrl}. 

We show that $\mathcal{E}_{\mathrm{dir}}(\bm{d})$ is also represented as a quadratic form of $\bm{d}$ in a similar manner to $\mathcal{E}_{\mathrm{uni}}(\bm{d})$. First, $w$ is expanded by spherical harmonic function $Y_n^m$ with order $n$ and degree $m$ around $\bm{r}_{\mathrm{S}}$ as
\begin{align}
 w(\theta,\phi) = \sum_{n,m} \tilde{w}_{n}^{m}(\bm{r}_{\mathrm{S}}) Y_{n}^{m} (\theta,\phi),
\label{eq:w_exp}
\end{align}
where $\tilde{w}_{n}^{m}(\bm{r}_{\mathrm{S}})$ is the expansion coefficients and $\sum_{n,m}:=\sum_{n=1}^{\infty}\sum_{m=-n}^n$. Besides, $u_{\mathrm{syn}}$ is expanded by spherical wavefunction for exterior field around $\bm{r}_{\mathrm{S}}$ for $\bm{r}=(r,\theta,\phi)\in\mathbb{R}^3\backslash\Omega_{\mathrm{S}}$ as
\begin{align}
 u_{\mathrm{syn}}(\bm{r}) = \sum_{n,m} \mathring{u}_{\mathrm{syn},n}^m(\bm{r}_{\mathrm{S}}) h_n(kr) Y_n^m(\theta,\phi),
\label{eq:usyn_exp}
\end{align}
where $\mathring{u}_{\mathrm{syn},n}^m(\bm{r}_{\mathrm{S}})$ is the expansion coefficients, and $h_n$ is the $n$th-order spherical Hankel function of the first kind. By substituting \eqref{eq:w_exp} and \eqref{eq:usyn_exp} into \eqref{eq:erp_dir}, $\mathcal{E}_{\mathrm{dir}}(\bm{d})$ is written as
\begin{align}
& \mathcal{E}_{\mathrm{dir}}(\bm{d}) = \notag\\
& \quad \frac{1}{2\rho c k} \mathrm{Re} \Big[ \mathrm{j} \sum_{\nu,\mu} \tilde{w}_{\nu}^{\mu}(\bm{r}_{\mathrm{S}}) \sum_{n,m} \sum_{n^\prime,m^{\prime}} \mathring{u}_{\mathrm{syn},n}^m(\bm{r}_{\mathrm{S}}) \mathring{u}_{\mathrm{syn},n^{\prime}}^{m^{\prime}}(\bm{r}_{\mathrm{S}})^{\ast} \notag\\
& \quad \cdot h_n(kR_{\mathrm{S}}) \frac{\partial}{\partial R_{\mathrm{S}}} h_{n^{\prime}}(kR_{\mathrm{S}})^{\ast} \notag\\
& \quad \cdot R_{\mathrm{S}}^2  \int_0^{2\pi} \mathrm{d}\phi \int_0^{\pi} \mathrm{d}\theta Y_n^m(\theta,\phi) Y_{n^{\prime}}^{m^{\prime}}(\theta,\phi)^{\ast} Y_{\nu}^{\mu}(\theta,\phi) \sin\theta \Big].
\label{eq:erp_dir1}
\end{align}
The integral of the triple product of spherical harmonic functions in \eqref{eq:erp_dir1} is rewritten as 
\begin{align}
 \int_0^{2\pi} \mathrm{d}\phi \int_0^{\pi} \mathrm{d}\theta Y_n^m(\theta,\phi) Y_{n^{\prime}}^{m^{\prime}}(\theta,\phi)^{\ast} Y_{\nu}^{\mu}(\theta,\phi) \sin\theta \notag\\
=\sum_{q=|n-\nu|}^{n+\nu} \mathcal{G}(n,m; \nu,\mu; q) \delta_{q,n^{\prime}} \delta_{m+\mu,m^{\prime}},
\end{align}
where $\mathcal{G}$ is referred to as \textit{Gaunt coefficients}~\cite{Martin:MultiScat}. Thus, $\mathcal{E}_{\mathrm{dir}}(\bm{d})$ is written as
\begin{align}
& \mathcal{E}_{\mathrm{dir}}(\bm{d}) \notag\\
&= - \frac{R_{\mathrm{S}}^2}{2\rho c} \mathrm{Im} \Big[  
\sum_{\nu,\mu} \sum_{n,m} \sum_{n^{\prime},m^{\prime}} \tilde{w}_{\nu,\mu}(\bm{r}_{\mathrm{S}}) \mathring{u}_{\mathrm{syn},n}^m(\bm{r}_{\mathrm{S}}) \mathring{u}_{\mathrm{syn},n^{\prime}}^{m^{\prime}}(\bm{r}_{\mathrm{S}})^{\ast} \notag\\
& \quad \cdot h_n(kR_{\mathrm{S}})h_{n^{\prime}}^{\prime}(kR_{\mathrm{S}})^{\ast}
\sum_{q=|n-\nu|}^{n+\nu} \mathcal{G}(n,m; \nu,\mu; q) \delta_{q,n^{\prime}} \delta_{m+\mu,m^{\prime}} \Big] \notag\\
&= - \frac{R_{\mathrm{S}}^2}{2\rho c} \mathrm{Im} \Big[  
\sum_{\nu,\mu} \sum_{n,m} \sum_{n^{\prime}=|n-\nu|}^{n+\nu} \tilde{w}_{\nu,\mu}(\bm{r}_{\mathrm{S}}) \mathring{u}_{\mathrm{syn},n}^m(\bm{r}_{\mathrm{S}}) \mathring{u}_{\mathrm{syn},n^{\prime}}^{m^{\prime}}(\bm{r}_{\mathrm{S}})^{\ast} \notag\\
& \quad \hspace{20pt} \cdot h_n(kR_{\mathrm{S}})h_{n^{\prime}}^{\prime}(kR_{\mathrm{S}})^{\ast}
 \mathcal{G}(n,m; \nu,\mu; n^{\prime}) \Big]. 
\label{eq:erp_dir2}
\end{align}

\begin{figure}[t]
    \begin{algorithm}[H]
        \caption{ADMM algorithm for amplitude matching with exterior radiation suppression}
        \label{alg1}
        \begin{algorithmic}[1] 
        \renewcommand{\algorithmicrequire}{\textbf{Input:}}
        \REQUIRE $\bm{d}^{(0)},\bm{\lambda}^{(0)}, \bm{a}_\mathrm{des}, \bm{G}, \bm{A},\gamma, \alpha, \xi$
        \STATE $t = 0$
        \WHILE{stopping criterion not satisfied}
        \STATE $\bm{h}^{(t)} = \bm{Gd}^{(t)} + \bm{\lambda}^{(t)}/\xi$
        \STATE $\bm{\theta}^{(t+1)} = \arg( \bm{h}^{(t)} )$
        \STATE $\bm{a}^{(t+1)} = (\xi|\bm{h}^{(t)}|+2\bm{a}_\mathrm{des}) / (\xi + 2)$
        \STATE $\bm{d}^{(t+1)} =  \left(\bm{G}^\mathsf{H}\bm{G} + (2\gamma / \xi) \bm{A} + (2\alpha / \xi) \bm{I}\right)^{-1}$\\
	 $\quad\quad\quad\quad \cdot\bm{G}^\mathsf{H}\left(\bm{a}^{(t+1)}\odot \mathrm{e}^{\mathrm{j}\bm{\theta}^{(t+1)}} - \bm{\lambda}^{(t)} / \xi \right)$
        \STATE $\bm{\lambda}^{(t+1)} = \bm{\lambda}^{(t)} + \xi(\bm{Gd}^{(t+1)} - \bm{a}^{(t+1)}\odot \mathrm{e}^{\mathrm{j}\bm{\theta}^{(t+1)}})$
        \STATE $t = t + 1$
        \ENDWHILE
        \STATE \textbf{return} $\bm{d}^{(t)}$ 
        \end{algorithmic}
    \end{algorithm}
\vspace{-5pt}
\end{figure}

Next, $\mathring{u}_{\mathrm{syn},n}^m(\bm{r}_{\mathrm{S}})$ is represented by a weighted sum of the expansion coefficients of $g_l(\bm{r})$ around $\bm{r}_{\mathrm{S}}$, i.e., $\mathring{g}_{l,n}^m(\bm{r}_{\mathrm{S}})$, with the weight of $d_l$~\cite{Ueno:IEEE_ACM_J_ASLP2019} as
\begin{align}
 \mathring{u}_{\mathrm{syn},n}^m(\bm{r}_{\mathrm{S}}) = \sum_{l=1}^L d_l \mathring{g}_{l,n}^m (\bm{r}_{\mathrm{S}}).
\label{eq:uyn_coef_sum_g_coef}
\end{align}
By using the translation operator $\hat{S}_{\nu,n}^{\mu,m}$, $\mathring{g}_{l,n}^m(\bm{r}_{\mathrm{S}})$ is represented by using the expansion coefficients around the secondary source position $\bm{r}_l$ as
\begin{align}
 \mathring{g}_{l,n}^m(\bm{r}_{\mathrm{S}}) = \sum_{\nu,\mu} \mathring{g}_{l,\nu}^{\mu}(\bm{r}_l) \hat{S}_{\nu,n}^{\mu,m}(\bm{r}_{\mathrm{S}}-\bm{r}_l),
\label{eq:g_coef_trans}
\end{align}
when $\|\bm{r}-\bm{r}_\mathrm{S}\| > \|\bm{r}_\mathrm{S} - \bm{r}_l\|$. Here, the translation operator $\hat{S}_{\nu,n}^{\mu,m}$ is defined as~\cite{Martin:MultiScat}
\begin{align}
 \hat{S}_{\nu,n}^{\mu,m}(\bm{r}) = &4\pi \mathrm{j}^{n-\nu} \sum_{q=|\nu-n|}^{\nu+n} \mathrm{j}^q (-1)^{\mu} j_q(kr) Y_{q}^{m-\mu}(\theta,\phi)^{\ast} \notag\\
& \quad \quad \cdot \mathcal{G}(\nu,\mu; n, -m; q).
\end{align}
By substituting \eqref{eq:uyn_coef_sum_g_coef} and \eqref{eq:g_coef_trans} into \eqref{eq:erp_dir2}, $\mathcal{E}_{\mathrm{dir}}(\bm{d})$ is represented by a quadratic form of $\bm{d}$ as
\begin{align}
 \mathcal{E}_{\mathrm{dir}}(\bm{d}) = \bm{d}^{\mathsf{H}} \bm{A} \bm{d}
\label{eq:erp_dir_qf}
\end{align}
with 
\begin{align}
 \bm{A} = -\frac{R_{\mathrm{S}}^2}{2\rho c} \mathrm{Im} \left[ \sum_{n,m} \tilde{w}_n^m(\bm{r}_{\mathrm{S}}) \bm{B}_n^m \right] 
%=\frac{\mathrm{j} R_{\mathrm{S}}^2}{4\rho c} \sum_{n,m} \left( \tilde{w}_n^m \bm{B}_n^m - \tilde{w}^{\ast} \bm{B}_n^{m \mathsf{H}} \right)  
\label{eq:erp_dir_qf_mat}
\end{align}
and 
\begin{align}
& (\bm{B}_n^m)_{l_1,l_2} = \sum_{p,q} \mathring{g}_{l_1,p}^q(\bm{r}_{l_1})^{\ast} \sum_{p^{\prime},q^{\prime}} \mathring{g}_{l_2,p^{\prime}}^{q^{\prime}}(\bm{r}_{l_2}) \notag\\
& \quad \quad \cdot \sum_{n,m}\sum_{n^{\prime}=|n-\nu|}^{n+\nu} \hat{S}_{p,n^{\prime}}^{q,(m+\mu)} (\bm{r}_{\mathrm{S}}-\bm{r}_{l_1})^{\ast} \hat{S}_{p^{\prime},n}^{q^{\prime},m} (\bm{r}_{\mathrm{S}}-\bm{r}_{l_2}) \notag\\
& \quad \quad \cdot h_n(kR_{\mathrm{S}}) h_{n^{\prime}}^{\prime}(kR_{\mathrm{S}})^{\ast} \mathcal{G}(n,m; \nu,\mu; n^{\prime}).
\label{eq:erp_dir_qf_matB}
\end{align}
Here, $\mathrm{Im}[\cdot]$ denotes the imaginary part of a complex number. It is necessary to truncate the expansion order to compute the summations in \eqref{eq:erp_dir_qf_mat} and \eqref{eq:erp_dir_qf_matB} in practice, which is a difference from the computation of $\bm{A}$ for $\mathcal{E}_{\mathrm{uni}}(\bm{d})$ based on an analytical representation given in~\cite{Ueno:ICASSP2018}. It is preferable that $w$ can be represented by low-order coefficients of $\tilde{w}_n^m$ to reduce the computational cost of $\bm{A}$.

\subsection{Application to Amplitude Matching}

As discussed in Sect.~\ref{sec:sfs}, $\mathcal{E}_{\mathrm{dir}}(\bm{d})$ formulated in \eqref{eq:erp_dir_qf} can be used in various sound field synthesis problems. As an example, we demonstrate an application of the directionally weighted exterior radiation suppression in amplitude matching. 

The control points are set at $\{\bm{r}_i\}_{i=1}^I$ over $\Omega$. The desired amplitude at the control points is denoted as $\bm{a}_{\mathrm{des}}\in\mathbb{R}_{\ge 0}^I$. The transfer function matrix between the secondary sources and control points is defined as $\bm{G}\in\mathbb{C}^{I \times L}$. The goal is to synthesize the desired amplitude $\bm{a}_{\mathrm{des}}$ at the control points while suppressing the exterior radiation power outside the secondary sources. Thus, the cost function is formulated as
\begin{align}
 \minimize_{\bm{d}\in\mathbb{C}^L} \left\| |\bm{Gd}| - \bm{a}_{\mathrm{des}}  \right\|^2 + \gamma \bm{d}^{\mathsf{H}} \bm{A} \bm{d} + \alpha \|\bm{d}\|^2,
 \label{eq:cost_am_ers}
\end{align}
where $|\cdot|$ represents the element-wise absolute value. Since the cost function in \eqref{eq:cost_am_ers} is neither convex nor differentiable, \eqref{eq:cost_am_ers} does not have a closed-form solution. In a similar manner to amplitude matching without exterior radiation suppression~\cite{Abe:IEEE_ACM_J_ASLP2023}, an algorithm based on alternating direction method of multipliers (ADMM) for solving \eqref{eq:cost_am_ers} can be derived, owing to the simple form of $\mathcal{E}_{\mathrm{dir}}(\bm{d})$, as in Algorithm~\ref{alg1}. Here, $\bm{\lambda}^{(t)}\in\mathbb{C}^I$ is the Lagrange multiplier, $\xi>0$ is the penalty parameter, and $t$ denotes the iteration index. Since \eqref{eq:cost_am_ers} is formulated for the single frequency, the filter to obtain the driving signal in the time domain can be unnecessarily long. It is also possible to apply the differential-norm penalty proposed in \cite{Abe:IEEE_ACM_J_ASLP2023} for $\mathcal{R}(\bm{d})$ to make the filter length small by inducing smoothness of $\bm{d}$ between frequency bins. 

\section{Experiments}

%\begin{figure*}[t]
% \begin{tabular}{ccc}
%  \begin{minipage}[t]{0.32\hsize}
%   \centering
%   \includegraphics[width=1.0\columnwidth]{fig/3D_SPKarr.eps}
%   \caption{Experimental setup. Blue dots and green cylinder represent loudspeakers and $\Omega$, respectively.}
%   \label{fig:exp_setup}
%  \end{minipage}&
%  \begin{minipage}[t]{0.32\hsize}
%   \centering
% \includegraphics[width=1.0\columnwidth]{fig/3D_AM-Prad_MSE.eps}
% \caption{MSE of AM, AM-Rad, and AM-Rad-Dir with respect to frequency.}
% \label{fig:mse_freq}
%  \end{minipage}&
%  \begin{minipage}[t]{0.32\hsize}
%   \centering
% \includegraphics[width=1.0\columnwidth]{fig/3D_AM-Prad_Prad.eps}
% \caption{$P_{\mathrm{rad}}$ of AM, AM-Rad, and AM-Rad-Dir with respect to frequency.}
% \label{fig:prad_freq}
%  \end{minipage}
% \end{tabular}
%\end{figure*}

\begin{figure}
    \centering
   \includegraphics[width=0.7\columnwidth]{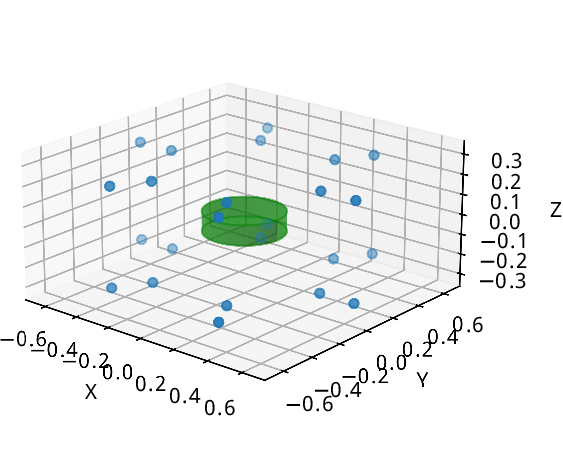}
   \caption{Experimental setup. Blue dots and green cylinder represent loudspeakers and target region $\Omega$, respectively.}
   \label{fig:exp_setup}
\end{figure}

We performed numerical experiments to evaluate the proposed exterior radiation suppression in amplitude matching. We compare amplitude matching (AM), AM with uniformly weighted exterior radiation suppression (AM-Rad), and AM with directionally weighted exterior radiation suppression (AM-Rad-Dir). 

Fig.~\ref{fig:exp_setup} shows the experimental setup. The target region $\Omega$ was a cylindrical region of radius $0.2~\mathrm{m}$ and height $0.1~\mathrm{m}$. 147 control points were uniformly distributed over $\Omega$ at intervals of $0.05~\mathrm{m}$. 6 loudspeakers were equiangularly placed on 4 circles of radii $0.453~\mathrm{m}$ and $0.653~\mathrm{m}$ at $z=\pm 0.2~\mathrm{m}$. Therefore, the total number of loudspeakers was 24. Each loudspeaker was assumed to be a point source. 

The directional weighting function $w$ was defined as
\begin{align}
 w(\theta,\phi) = 1 + \cos\phi \sin \theta,
\end{align}
which allows exterior radiation in the negative direction of the $x$-axis and suppresses it in the positive direction. The matrix $\bm{A}$ in \eqref{eq:erp_dir_qf_mat} was computed by truncating the expansion coefficients up to $\lceil k R_{\mathrm{S}} \rceil$ with $R_{\mathrm{S}}=0.8~\mathrm{m}$. The desired amplitude at the control points $\bm{a}_{\mathrm{des}}$ was set to $1$, i.e., uniform amplitude over $\Omega$. The time-domain driving signal was obtained by using the differential-norm penalty. The parameter for the differential-norm penalty was empirically set at $10$. The balancing parameter $\gamma$ was set at $2.0 \times 10^4$, which was the largest value so that the interior synthesis accuracy was kept high below $1000~\mathrm{Hz}$.  

%We set the hyperparameter $\rho = 1$, and the regularization parameter $\alpha = 10$ empirically. The balanced parameter $\gamma$ = 20000 is the largest value which flattens the exterior radiation energy at each frequency. (If $\gamma$ is too large, radiation energy at high frequency is extremely suppressed.)

As evaluation measures, we define the mean square error of the amplitude synthesized at the control points for each frequency ($\mathrm{MSE}(\omega)$) and exterior radiation power in the suppression directions ($P_{\mathrm{rad}}$) as
\begin{align}
 & \mathrm{MSE}(\omega) = \frac{1}{M} \left\| \bm{a}_{\mathrm{syn}}(\omega) - \bm{a}_{\mathrm{des}}(\omega) \right\|^2 \\
 & P_{\mathrm{rad}}(\omega) = \int_{\phi_1}^{\phi_2} \mathrm{d}\phi \int_{\theta_1}^{\theta_2} \mathrm{d}\phi I_{r}(\theta,\phi) r^2 \sin\theta, 
\end{align}
where the range of the suppression direction is set by $(\theta_1,\theta_2)=(0,\pi)$ and $(\phi_1,\phi_2)=(-\pi/2,\pi/2)$, i.e., the directions of $w>1$. 

\begin{figure}
    \centering
    \includegraphics[width=0.9\columnwidth]{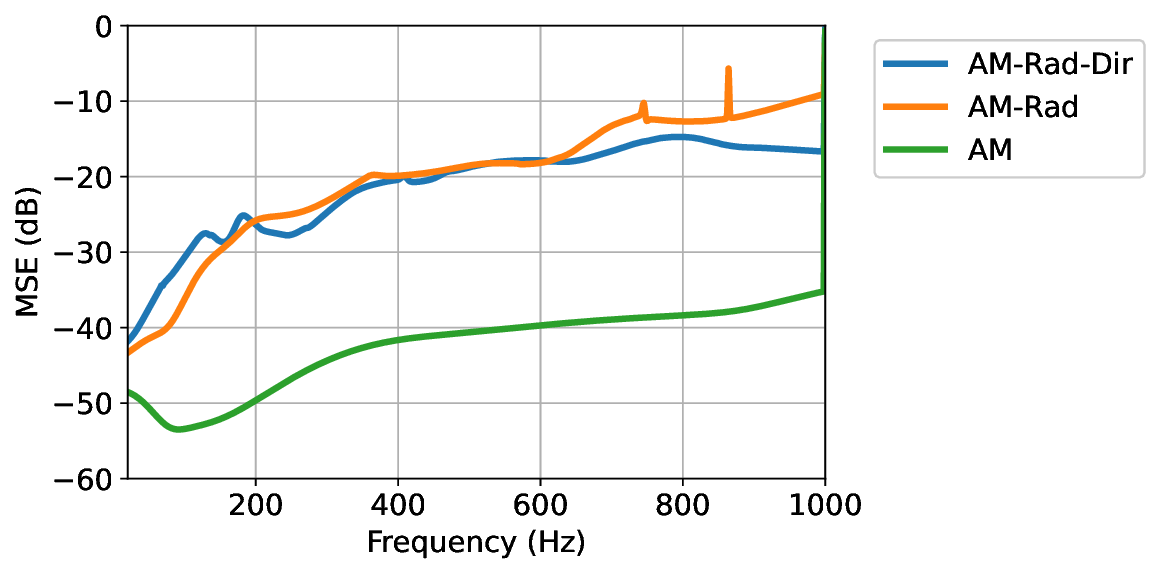}
 \caption{MSE of AM, AM-Rad, and AM-Rad-Dir with respect to frequency.}
 \label{fig:mse_freq}
\end{figure}

\begin{figure}
    \centering
 \includegraphics[width=0.9\columnwidth]{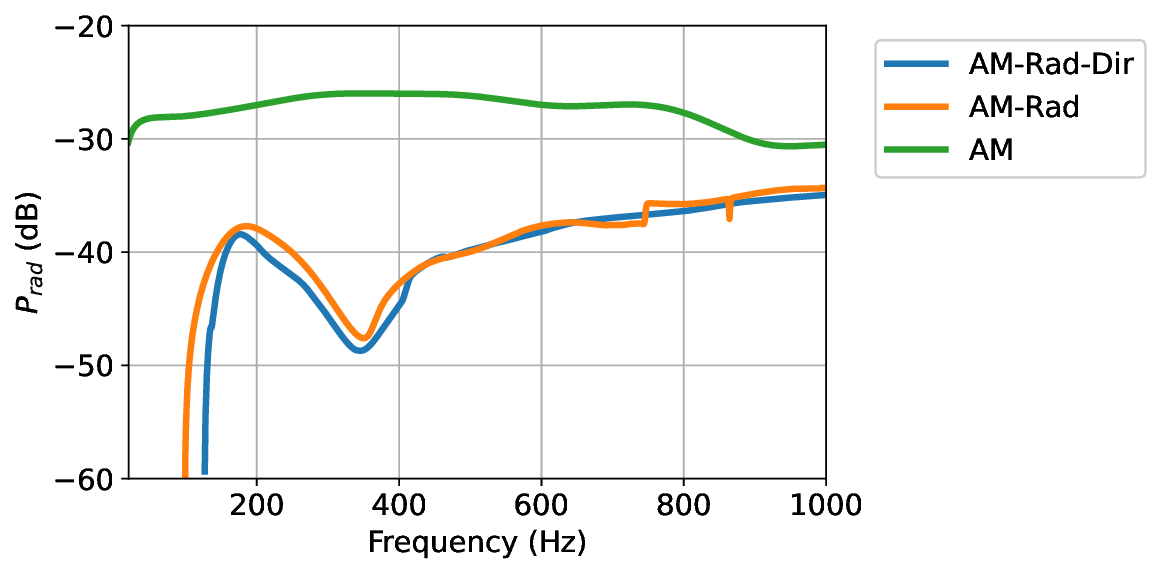}
 \caption{$P_{\mathrm{rad}}$ of AM, AM-Rad, and AM-Rad-Dir with respect to frequency.}
 \label{fig:prad_freq}
\end{figure}

$\mathrm{MSE}(\omega)$ and $P_{\mathrm{rad}}(\omega)$ are plotted with respect to frequency in Figs.~\ref{fig:mse_freq} and \ref{fig:prad_freq}. Although $\mathrm{MSE}(\omega)$ of AM-Rad and AM-Rad-Dir was higher than that of AM, their $P_{\mathrm{rad}}(\omega)$ was significantly smaller than that of AM. Comparing AM-Rad and AM-Rad-Dir, $P_{\mathrm{rad}}(\omega)$ of AM-Rad-Dir was smaller than that of AM-Rad below $400~\mathrm{Hz}$, owing to the directional weighting. Furthermore, $\mathrm{MSE}(\omega)$ of AM-Rad-Dir was smaller than that of AM-Rad above $600~\mathrm{Hz}$, which will be owing to the relaxation of the constraint on the exterior radiation power by directional weighting. 

\begin{figure}[t]
 \centering
 \subfloat[AM]{\includegraphics[width=0.49\columnwidth]{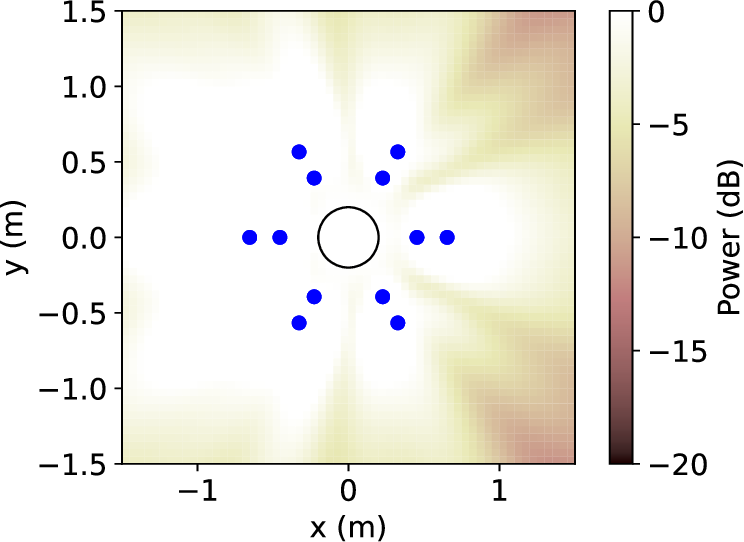}}\\
 \subfloat[AM-Rad]{\includegraphics[width=0.49\columnwidth]{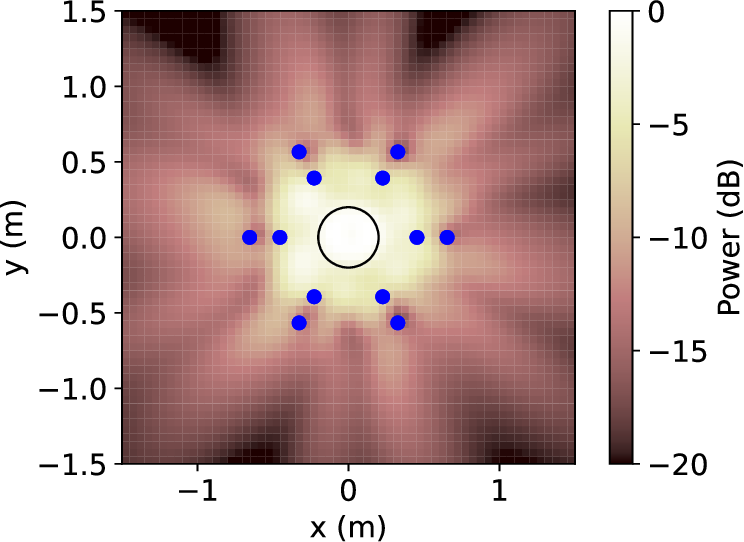}}%
 \hspace{2pt} \subfloat[AM-Rad-Dir]{\includegraphics[width=0.49\columnwidth]{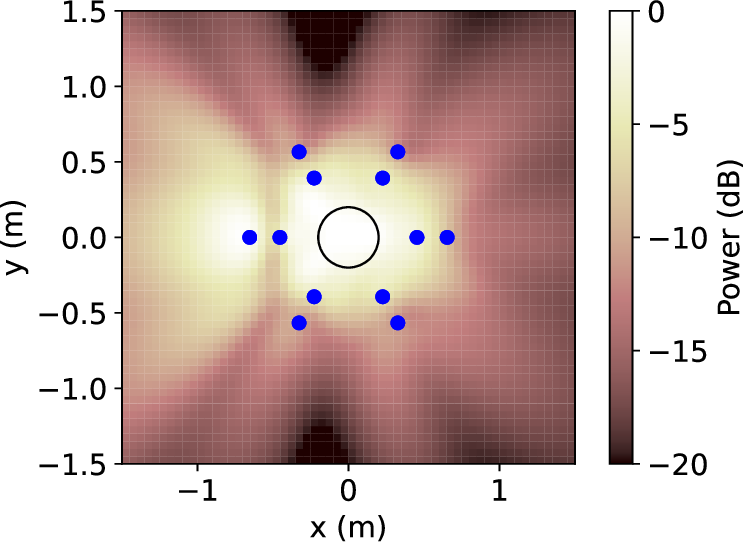}}
 \caption{Power distribution of synthesized sound field on $x$-$y$-plane at $z=0$.}
\label{fig:eval_dist}
\end{figure}

Fig.~\ref{fig:eval_dist} shows the power distribution of the synthesized sound field up to $1000~\mathrm{Hz}$ on $x$-$y$-plane at $z=0$. The exterior radiation of AM was particularly larger than that of AM-Rad and AM-Rad-Dir because the exterior radiation suppression is not taken into consideration in AM. In AM-Rad, the exterior radiation was suppressed in all directions. Because of the design of the directional weighting function $w$, the exterior radiation in the negative $x$ direction was larger than that in the positive $x$ direction in AM-Rad-Dir.

\section{Conclusion}

We proposed a method for synthesizing the desired sound field while suppressing the exterior radiation power with directional weighting. The proposed directionally weighted exterior radiation suppression enables relaxing the constraint on the exterior radiation compared with uniformly weighted one, thus the influence on the interior synthesis accuracy owing to the constraint on the exterior radiation can be mitigated. Since the directionally weighted exterior radiation power is represented by a quadratic form of the driving signal of the secondary sources, it is simply applicable to various sound field synthesis methods. In the numerical experiments using amplitude matching, the exterior radiation was significantly suppressed in the prioritized directions while maintaining the interior synthesis accuracy high. 

\section{Acknowledgment}

This work was supported by JST FOREST Program Grant Number JPMJFR216M and JSPS KAKENHI Grant Number 22H03608.

\vfill
%\newpage

\bibliographystyle{IEEEbib_mod}
\bibliography{str_def_abrv,koyama_en,refs}

\end{document}